%% file: sqrt-part1.tex
\documentclass[12pt, nofootinbib]{article}

\include{jm-tex-macros-public}

\include{no-revtex}
\usepackage{braket}

\newcommand{\beq}{\begin{equation}}
\newcommand{\eeq}{\end{equation}}

\newcommand{\MZ}{\mathcal{Z}}

\begin{document}

\begin{titlepage}

\title{Renormalization Group Circuits for Gapless States}


\author{Brian Swingle${}^a$, John McGreevy${}^b$ and Shenglong Xu${}^b$}

\address{${}^a$ Department of Physics, Stanford University, Palo Alto, CA 94305, USA}

\address{${}^b$ Department of Physics, University of California at San Diego, La Jolla, CA 92093, USA}

\begin{abstract}
We show that a large class of gapless states are renormalization group fixed points 
in the sense that they can be grown 
scale by scale using local unitaries.
This class of examples includes some theories with dynamical exponent different from one, but does not include conformal field theories. The key property of the states we consider is that the ground state wavefunction is related to the statistical weight of a local statistical model. We give several examples of our construction in the context of Ising magnetism.
\end{abstract}

\vfill

\today

\end{titlepage}

\vfill\eject

\setcounter{tocdepth}{2}    
\tableofcontents

\vfill\eject

\section{Introduction}

In this work we are interested in the entanglement structure of quantum critical points. These are systems where, in the thermodynamic limit, the spectrum of the Hamiltonian has no energy gap and the system displays scale invariant physics. Much effort has been expended to quantify entanglement in quantum critical points using entanglement entropy, but such a characterization is only the first step towards a complete understanding of the entanglement structure of these states of matter. A more refined characterization of the pattern of entanglement is provided by a quantum circuit which produces the state of interest from a product state---in essence we seek a set of instructions for building up the entanglement in the state from elementary ingredients.

Based on the scale invariance of the physics we expect the entanglement to be organized in a scale invariant way. This expectation is encoded in various circuit networks which have been conjectured to be capable of approximating well the ground state wavefunction of scale invariant states. These networks include MERA 
\cite{mera} and branching MERA \cite{PhysRevLett.112.240502} 
and the more general notion of $s$ sourcery \cite{2014arXiv1407.8203}. However, while it is physically quite reasonable to conjecture that such circuits can well approximate the ground state, the actual evidence that this is true is scarce. In one dimension there is excellent numerical evidence that MERA well approximates the states of simple conformal field theories \cite{2011arXiv1109.5334E}. In higher dimensions the primary evidence comes from free field theory 
\cite{PhysRevB.81.235102, PhysRevLett.113.010401, 2015WhiteFreeFermions}
and from holographic models \cite{Swingle:2009bg}; in the latter case it was proposed that the geometry of the quantum circuit was related to the emergent holographic geometry. Moreover, we have very little information about the degree of approximation involved in using such a scale invariant circuit with a fixed bond dimension
(the range of the indices of the tensors constituting the circuit).

The purpose of this paper is to partially remedy the above deficiencies by producing renormalization group circuits for certain wavefunctions supporting scale invariant physics. We are motivated by a desire to make progress -- rigorous if possible -- showing that gapless quantum phases and quantum critical points can be accurately captured, at the level of wavefunctions, within a renormalization group (RG) framework,
along the lines of \cite{2014arXiv1407.8203}. 
Specifically, we will show that a large class of such states are $s=1$ RG fixed points 
in the sense defined in \cite{2014arXiv1407.8203}.
The meta-motivation is twofold: (1) to understand the entanglement structure of quantum matter, {\it e.g.}~for simulation purposes, and (2) to further the Einstein from qubits story of emergent gravity \cite{Swingle:2009bg, Swingle:2012wq, 2014JHEP...03..051F, Hartman:2013qma, 2014arXiv1405.2933S}.

The class of wavefunctions we are interested in are those that arise from the statistical weight of a classical statistical model\footnote{States of this form have a long history.
The earliest references of which we are aware arise in the
context of studies of kinetics of the Ising model
\cite{1979JSP....21..289K, 1981ZPhyB..43..241P}, and more recent work
includes \cite{Rokhsar:1988zz, PhysRevB.64.144416, 2002PhRvB..65b4504M, Ardonne:2003wa, Horava:2009uw, 2016MonthusSQRT}.}. Let us work with spins $Z_r$ ($r$ labels sites of a lattice) for concreteness (generalizations are obviously possible). Given a classical Hamiltonian $h(Z)$ we form the quantum wavefunction
\beq
\psi(Z;h,\beta) = \sqrt{   \frac{e^{- \beta h(Z)}}
{\MZ}}
\eeq
where $\beta$ is a temperature we choose and $\MZ$ is the classical partition function of the statistical model determined by $h$ and $\beta$.\footnote{A construction of a PEPS representation of such states
was made in \cite{2006PhRvL..96v0601V}.
In general a PEPS is not efficiently contractible however the technology we use to produce our circuits also permit these particular PEPS networks to be contracted. It would be quite interesting to understand if this is a more general connection - that a PEPS inherits contractibility from the existence of an RG circuit.} We call such a state a square root state.

\subsection{Correlation structure of $\psi$}

$\psi$ is normalized according to
\beq
\sum_{Z} |\psi(Z)|^2 = 1
\eeq
which is the statement that the classical probabilities add up to one. For classical correlators, \eg~$\langle Z_r Z_{r'}\rangle$, the quantum correlation function is identical to the correlation function in the classical statistical model. This is because the expectation value,
\beq
\sum_Z |\psi(Z)|^2 Z_r Z_{r'} = \sum_Z \frac{e^{-\beta h(Z)}}{\MZ} Z_r Z_{r'},
\eeq
is manifestly given by the classical correlation function. This statement is true for any correlator consisting entirely of classical variables, {\it i.e.} variables in which $h$ is diagonal. Non-classical correlations, {\it e.g.} $\langle X_r X_{r'} \rangle$, are more complicated.

If $\beta$ corresponds to a critical temperature of the classical model $h$, then the classical correlations of $\psi$ will be power law. Hence $\psi$ necessarily describes a gapless state of matter, since gapped phases always have short-range correlations if the quantum Hamiltonian is local\footnote{Ordered groundstates of local Hamiltonians can have correlations which 
do not fall off with distance; such a system is gapless in the sense that 
the groundstate is degenerate in the thermodynamic limit.  We will focus rather on examples
where the correlations fall off with a nonzero power of the separation.
}
 (we will exhibit local quantum Hamiltonians whose groundstate is $\psi$ in examples). On the other hand, the wavefunction is relatively simple and accessible, so this class of quantum states is an attractive setting to explore wavefunction RG for gapless states.

\subsection{Entanglement structure of $\psi$ }

It is easy to see that $\psi$ (a square root state build from a
classical statistical model with a short-ranged Hamiltonian) has no entanglement between distant regions even when it hosts long-range correlations. Let $|\psi\rangle = \sum_Z \psi(Z)|Z\rangle$ be the normalized quantum state and let $ABC$ be a partition of the entire system such that $B$ separates $A$ from $C$, e.g. $B$ is an annulus, $A$ is the interior disk, and $C$ is the rest of the world. Let $\Pi(Z_B)$ be a projector onto a state of definite $Z_B$ (the spins in $B$). Then we have
\beq
\Pi(Z_B) |\psi\rangle =\sqrt{p(Z_B)} |\psi_A(Z_B)\rangle |Z_B\rangle |\psi_C(Z_B)\rangle,
\eeq
or in words, fixing the state of region $B$ causes the state of the whole system to factorize. This implies that the state $\rho_{AC}$ is separable. In detail, we have
\beq
\rho_{AC} = \tr_B(\rho_{ABC}) = \sum_{Z_B} \tr_B(\Pi(Z_B)\rho_{ABC}\Pi(Z_B)) = \sum_{Z_B} p(Z_B) \rho_A(Z_B)\rho_C(Z_B)
\eeq
which is manifestly an incoherent mixture. Thus $A$ and $C$ share no entanglement and the state $\rho_{AC}$ cannot be used to violate a Bell inequality despite the presence of long-range correlations.

By contrast, in a conformal field theory there is always some entanglement between $A$ and $C$ present in the ground state. This tells us that conformal field theories are not captured in the present construction.

Nevertheless, such square root states can still be long-range entangled. This must be true because, as we show later, topologically ordered states can sometimes be written as square root states. So while distant regions in the state cannot be used to violate a Bell inequality, the state is long-range entangled in the sense that it requires a high depth circuit to produce from an unentangled starting point. Indeed, the main purpose of this work is to exhibit such RG circuits.

\subsection{Some examples}

\noindent
\textit{Paramagnet -} The simplest possible example is where $h$ is a paramagnet: $h(Z) = - \sum_r Z_r$. In this case the resulting quantum wavefunction is a product state with no entanglement, but the onsite states are in a superposition of $\uparrow: Z=1\uparrow$ and $\downarrow: Z=-1$. The ratio of probabilities are the same as those of the classical model. The PEPS representation of such a state is trivial.

\noindent
\textit{Ferromagnet -} Another simple example occurs for the Ising model in 1d: $h(Z) = - \sum_r Z_r Z_{r+1}$. If we take the temperature $\beta = \infty$, then the resulting quantum state only has support on two configurations, all up and all down. Hence the quantum state for $L$ spins is a cat state,
\beq
|\text{cat}_L\rangle = \frac{1}{\sqrt{2}}(|0...0\rangle + |1...1\rangle ).
\eeq
In this equation we have switched to computational notation; $0$ corresponds to $\uparrow$ and $1$ corresponds to $\downarrow$, equivalently $Z = 1 - 2 x$ where $x=0$ or $x=1$ (not to be confused with $X$, the Pauli matrix). This state has a matrix product (MPS \cite{Fannes:1990ur, murg2008, 2011AnPhy.326...96S, Orus:2013kga}) representation 
\be\label{eq:MPS-0} 
\ket{h} =  \tr \prod_{i} A^{\sigma_i}    \ket{\{s_i\}}\ee
in which the matrices may be taken to be $A^0 = |0\rangle \langle 0|$ and $A^1 = |1\rangle \langle 1|$.

A simple protocol for producing the cat state is obtained by copying in the classical basis. Start with the cat state on $L$ sites, $|\text{cat}_L\rangle$. To make $|\text{cat}_{2L}\rangle$ introduce $L$ spins in the state $|0\rangle$. Intercalate the unentangled spins into the entangled spins to form a chain of length $2L$ where every other spin is unentangled. Now apply $L$ copy gates (CNOT will work) to each pair of one entangled and one unentangled spin. The copy gate performs $|00\rangle \rightarrow |00\rangle$ and $|10\rangle \rightarrow |11\rangle$. Then since the control bits are perfectly correlated it follows that the resulting state is $|\text{cat}_{2L}\rangle$.

\noindent
\textit{General Ising magnet -} For the bulk of the paper we focus on square root states derived from classical Ising magnetism in various dimensions. In 1d we will describe an exact RG circuit which produces the ground state while in 2d we will develop an systematic scheme to produce an approximate circuit. We will give bounds on the error of approximation using properties of the Ising magnet. The techniques we describe for the 1d and 2d Ising magnets generalize to more complicated classical statistical models.

\subsection{Precise problem}

What precisely would we like to do? Following the MERA and s-source RG story, we would like to produce a constant depth circuit that
maps the quantum state (plus initially unentangled degrees of freedom) at linear size $L$ to the quantum state at linear size $2L$. Such a circuit succinctly captures our intuition that gapless states describe some kind of RG fixed point. We would like to make this intuition sharp and eventually tackle CFTs and even more general gapless models. A first step is to understand the long-range states arising as square root states.

The problem can be decomposed into three parts.
\begin{description}
\item{Module 1.} The first part is purely classical:
Given a statistical lattice model,
identify a real-space RG scheme which
produces a model of the same form on a larger lattice.
This is particularly interesting for fixed-point values of the couplings.
This involves (at least) two sub-modules:
(1) The first is a geometric question of a re-wiring procedure on the lattice
which produces the larger lattice.
(2) The second is a map on the couplings for a specific model on said lattice.
\item{Module 2.}
Now consider the associated quantum state on a lattice of linear size $L$, $ \ket{h_L}$.
Turn the above RG map into a
unitary transformation which takes the given state
(plus ancillas) to the state on a larger geometry (perhaps plus other ancillas):
$$ \UU \ket{h_L} \otimes \ket{0...} = \ket{h_{2L}} \otimes \ket{0'...} . $$
Note that it may be necessary to increase the size of the on-site Hilbert space
(we will sometimes call this the `bond dimension'),
or make it infinite, to accomplish such an exact map.
\item{Module 3.}
Bound the error made by truncating the bond dimension in the previous step,
as a function of the bond dimension,
and as a function of the range of the classical Hamiltonian $h$.
\end{description}

The payoff of this construction is an efficiently-contractible representation of
the groundstate. Here is a brief guide to the results in this paper.
In \S\ref{sec:1d-sqrt}, we make an RG circuit
for the square root state associated to the Ising chain;
although this is a degenerate case, it is an instructive warmup.
In \S\ref{sec:ising-sqrt} we implement these steps
for the case of the quantum square root state for the general Ising model, focusing
on two spatial dimensions.
This model wavefunction exhibits several phases separated by quantum phase transitions.
In \S\ref{sec:bound-on-z} we provide a bound on the dynamical exponent
in the quantum critical point associated with the Onsager transition.
In \S\ref{sec:unitaries} we provide more details about the local unitaries for this state.
In \S\ref{sec:other} we discuss generalizations to other square root states,
including cases where the classical model is not short-ranged.

\section{1d Ising square root state}
\label{sec:1d-sqrt}

In this section we make a quantum circuit construction of the square root state associated with the Ising chain.
Though the state in question always has a finite correlation and only short-range entanglement, the correlation length can become exponentially large as a function of $\beta J$. Hence it is a natural toy model to begin with. Furthermore, the construction gives a clear demonstration of the capability of the RG circuit to compute useful information, such as correlation functions.

In the 1d case, the Ising square root state is:
\begin{equation}
\ket{h}=\frac{1}{\sqrt{\mathcal{Z}}}\sum\limits_{\{s\}}e^{+\frac{\beta J}{2}\sum\limits_{i}s_is_{i+1}}\ket{\{s_i\}}
\label{eq:1d_is_st}
\end{equation}
where $\mathcal{Z}$ is the partition function for 1D classical Ising model.
This state is a rank 2 matrix product state
\be\label{eq:MPS} 
\ket{h} = \(\frac{1}{\sqrt{\mathcal{Z}}} \)^N \tr \prod_{i} A^{\sigma_i}    \ket{\{s_i\}}\ee
 with
\begin{equation}
A^\sigma=
\begin{pmatrix}
\cosh(\frac{\beta J}{2}) & \sigma \sqrt{\cosh(\frac{\beta J}{2})\sinh(\frac{\beta J}{2}}) \\
\sigma \sqrt{\cosh(\frac{\beta J}{2})\sinh(\frac{\beta J}{2})}  &\sinh(\frac{\beta J}{2})\\
\end{pmatrix}~~.
\end{equation}

A parent Hamiltonian, of which this state is the ground state,  is
\begin{equation}
H=\sum\limits_i(-X_i+e^{-\beta J Z_i(Z_{i-1}+Z_{i+1})})
\label{eq:1d_is_ham}
\end{equation}

The physics of this model is simple. The system is always in a paramagnetic phase where $\langle Z \rangle = 0$, but as $\beta$ gets large the system develops increasingly long-ranged correlation without ever truly reaching a critical point. This is because the wavefunction is based on the 1d statistical Ising model which displays no phase transition and never supports power law correlations in the thermodynamic limit.
More directly from
\eqref{eq:1d_is_ham}, 
this is because
\eqref{eq:1d_is_ham}
contains antiferromagnetic interactions 
between nearest neighbors and next-nearest neighbors
of equal strength, and so is highly frustrated.

To verify these claims one can compute correlation functions
of local operators in the ground state via transfer matrix method. In this model the transfer matrix is defined as
\begin{equation}
T=e^{\beta J}I+e^{-\beta J}X;
\end{equation}
$T$ is diagonalized by the unitary matrix $u=\frac{1}{\sqrt{2}}(Z+X)$, so that:
\begin{equation}
uTu=2
\begin{pmatrix}
\cosh(\beta J) & 0\\
0                   & \sinh(\beta J)\\
\end{pmatrix}
\end{equation}
Therefore the partition function is
\begin{equation}
\mathcal{Z}=2^N(\cosh(\beta J)^N+\sinh(\beta J)^N)
\end{equation}

The $ZZ$ correlation function is
\begin{equation}
\begin{aligned}
C^{zz}(r)&=\bra{h} Z(r)Z(0) \ket{h} \\
              &=\frac{\tanh(\beta J)^{N-r}+\tanh(\beta J)^{r}}{1+\tanh(\beta J)^N}.
\end{aligned}
\end{equation}
The $XX$ correlation function can be computed using the matrix product representation
\eqref{eq:MPS}, and
 is
\begin{equation}
C^{xx}(r)=\bra{h}X(r)X(0)\ket{h}=\frac{\cosh(\beta J)^{-4 + N}}{\cosh(\beta J)^N + \sinh(\beta J)^N},
\end{equation}
independent of the separation, and disconnected: $\bra{h}X(r)X(0)\ket{h} =\(\bra{h}X(0)\ket{h}\)^2 $.

\subsection{RG circuit}

The 1d Ising square root state which we just introduced provides a simple exactly solvable example of an RG circuit. This is because the 1d statistical Ising model enjoys an \textit{exact} real space renormalization group, in the sense that one can trace out half of the spin degrees of  freedom in the partition function and obtain a new partition function with the same form but renormalized temperature. This procedure can be illustrated using three spins as follows:
\begin{equation}
\begin{aligned}
\sum\limits_{s_i}&e^{\beta J (s_{i-1}s_i +s_i s_{i+1})}\\
&=e^{\beta J (s_{i-1}+s_{i+1})}+e^{-\beta J (s_{i-1}+s_{i+1})}\\
&=2\sqrt{\cosh(2\beta J)}e^{\frac{1}{2}\ln \cosh(2\beta J)s_{i-1}s_{i+1}}
\end{aligned}
\end{equation}
Therefore the renormalized temperature is:
\begin{equation}
\tilde{\beta}J=\frac{1}{2}\ln \cosh(2\beta J)~.
\label{eq:t_flow}
\end{equation}
\begin{figure}
\begin{center}
\includegraphics[height=0.5\columnwidth, width=0.5\columnwidth]
{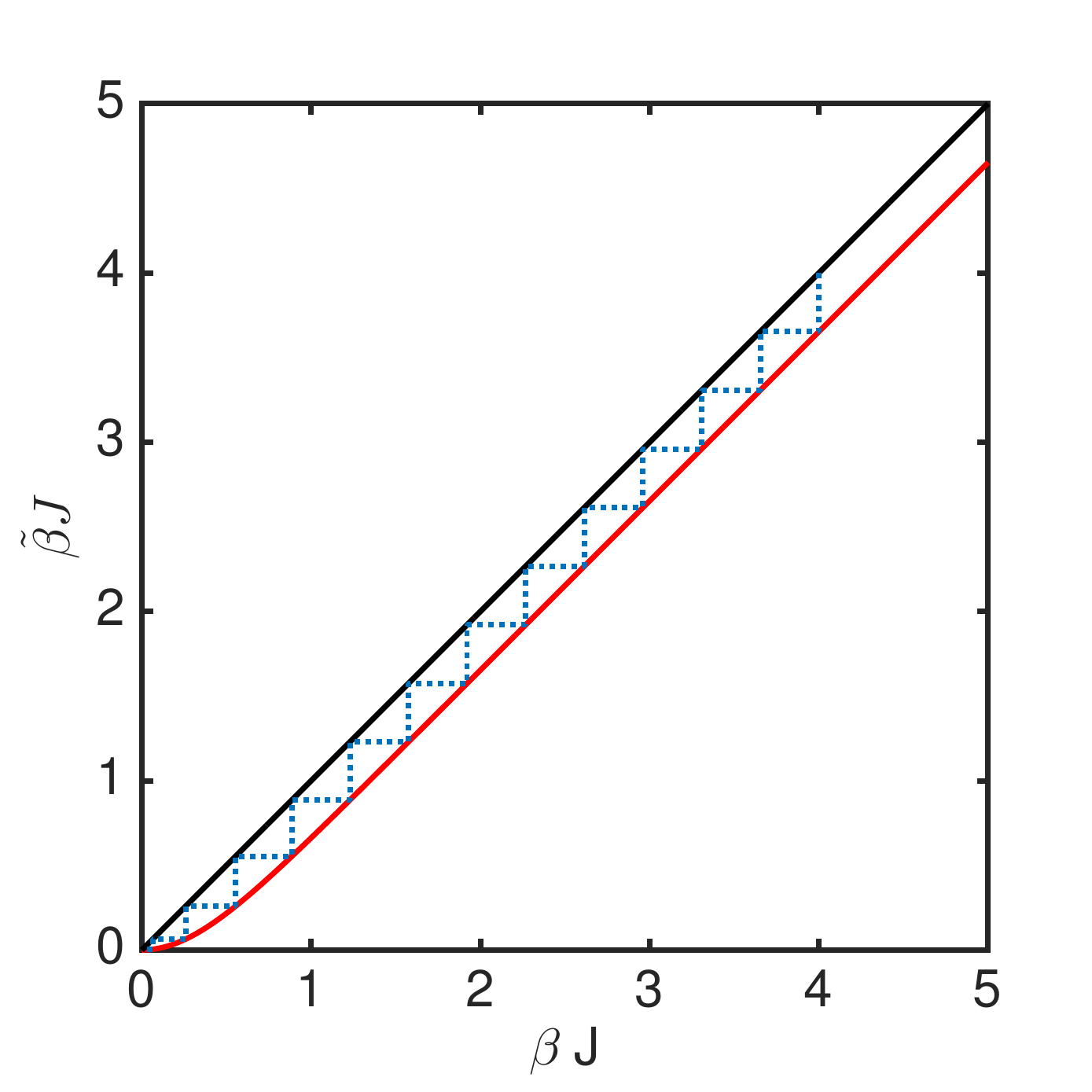}
\end{center}
\caption{temperature flow of real space RG for 1D classical Ising model}
\label{fig:t_flow}
\end{figure}
\noindent
There are two fixed points: the unstable low temperature fixed point and the stable high temperature fixed point. Therefore, under the RG flow, the classical Ising model, if not completely ordered, eventually flows to a completely disordered phase.

\begin{figure}
\begin{center}
\includegraphics[height=0.7\columnwidth, width=0.7\columnwidth]
{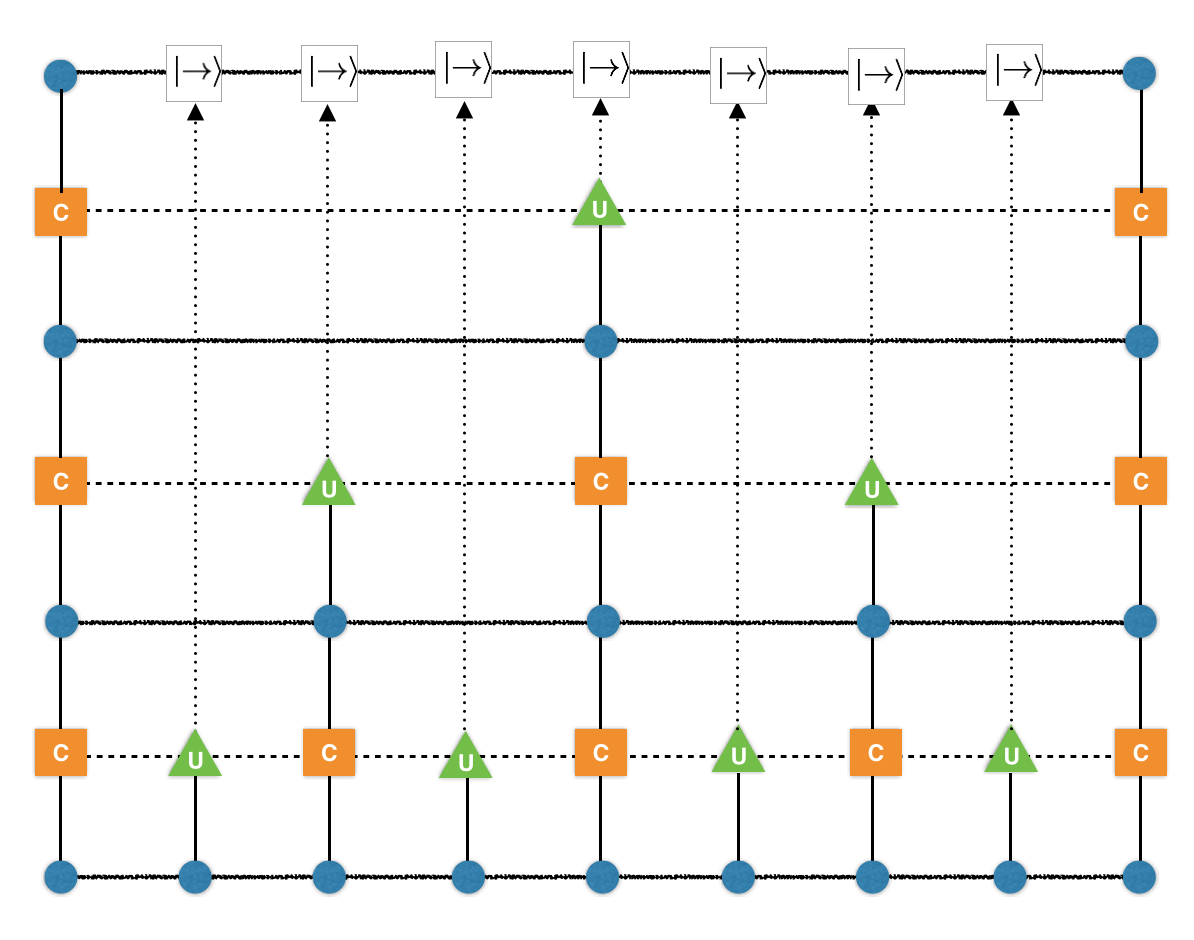}
\end{center}
\caption{Sketch of the RG circuit for the quantum Ising chain.}
\label{fig:circuit}
\end{figure}

\begin{figure}
\includegraphics[height=0.4\columnwidth, width=1.0\columnwidth]
{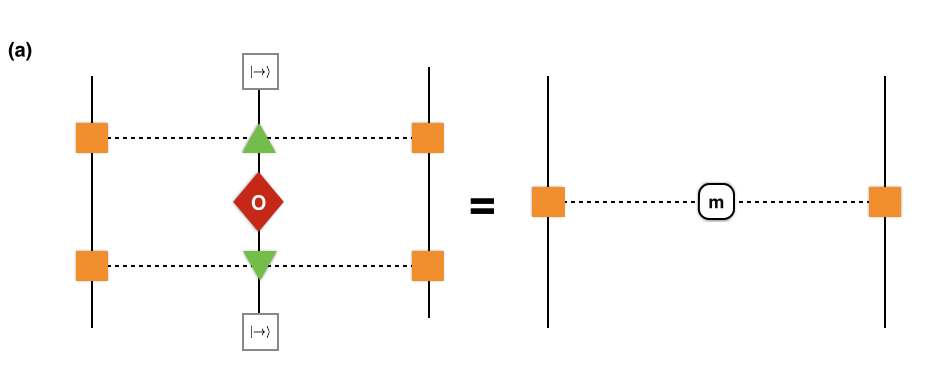}
\includegraphics[height=0.9\columnwidth, width=1.0\columnwidth]
{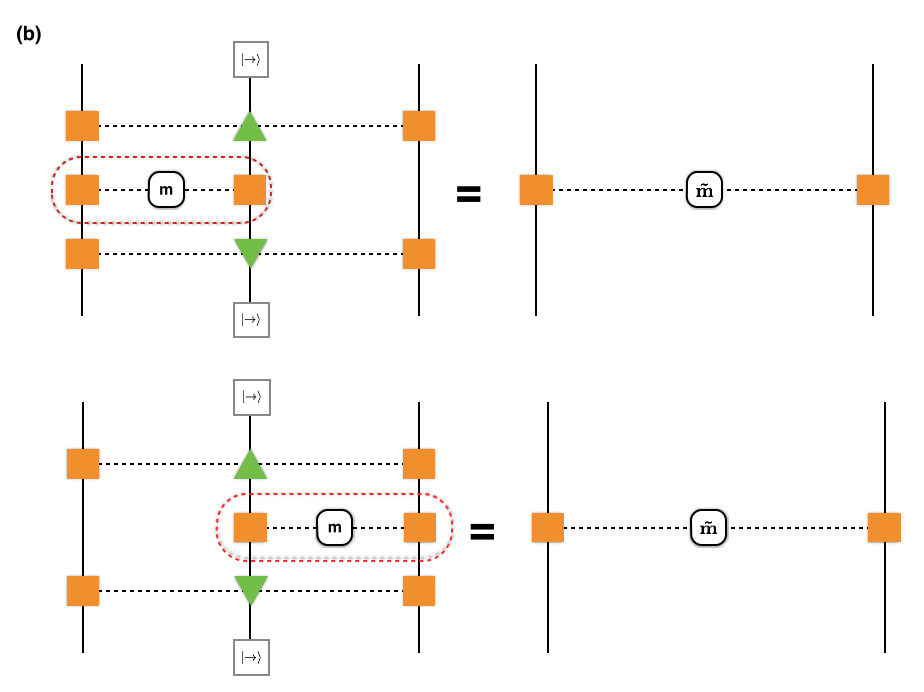}
\caption{Transformation of single site operator under unitary-RG transformation.}
\label{fig:single_site}
\end{figure}

Now let us explore the resulting RG circuit in the quantum theory. We first discuss the RG transformation of the state (Eq. \eqref{eq:1d_is_st}) then the Hamiltonian (Eq. \eqref{eq:1d_is_ham}). In the state, a single site spin state is completely determined once its neighboring spin states are fixed, and we have the freedom to apply a local unitary transformation to transform this spin state into an arbitrary state we desire. Consider a subset of three spins in the whole chain with the left and right spins fixed:
\begin{equation}
\ket{\psi_i}=\sum\limits_{s_i}e^{\frac{\beta J}{2}(s_{i-1}s_i+s_is_{i+1})}\ket{s_{i-1}s_is_{i+1}}.
\end{equation}
There exists a unitary transformation $u_i$ which disentangles the middle qubit:
\begin{equation}
u_i\ket{\psi_i}=\sqrt{2}\cosh^{\frac{1}{4}}(2\beta J)e^{\frac{\tilde{\beta}J}{2}s_{i-1}s_{i+1}}\ket{s_{i-1}s_{i+1}}\otimes\ket{\rightarrow_i}
\end{equation}
The explicit form of the unitary matrix $u_i$ is:
\begin{equation}
u_i=\Ione
+\delta_{s_{i-1}s_{i+1}}
\begin{pmatrix}
\frac{\cosh(\beta J)}{\sqrt{\cosh(2\beta J)}}-1 &   -s_{i-1}\frac{\sinh(\beta J)}{\sqrt{\cosh(2\beta J)}}\\
s_{i-1}\frac{\sinh(\beta J)}{\sqrt{\cosh(2\beta J)}} &\frac{\cosh(\beta J)}{\sqrt{\cosh(2\beta J)}}-1 &
\end{pmatrix}
\end{equation}
Then $U \equiv \prod\limits_{i \in odd}u_i$ puts all spins on odd sites into a product state of spin right and convert all even sites spins into a new Ising square root state with the renormalized temperature $\tilde{\beta}$:
\begin{equation}
U\ket{h}=\prod\limits_{i \in \text{odd}}u_i \ket{h(\beta)}=\ket{h\(\tilde{\beta}, \text{even}\)}\otimes\prod\limits_{i\in \text{odd}}\ket{\rightarrow_i}~.
\end{equation}
The $u_i$ commute from each other, therefore the product of them is also unitary. After this unitary transformation, the even site spins and odd site spins are completely disentangled with each other. Furthermore, the odd site spins are in a product state. When we repeatedly apply the above RG circuit, $\beta$ for the new square root state approaches zero and the unitary transformation approaches the identity.
As a result,
we obtain a product of all spin-right states, which is the stable fixed point of this unitary RG transformation, depicted in Fig. \ref{fig:circuit} as a circuit.
Expectation values of any operators $\CO$ in the ground state can be written in the following form:
\begin{equation}
\bra{h}
\mathcal{O}\ket{h}=\bra{X}U^{\nd} \mathcal{O}U^\dagger \ket{X}
\end{equation}
where $\ket{X}$ is just the product of right spins. Now we demonstrate how to use our RG circuit to compute the above quantity. In the case that $\mathcal{O}$ has a support of a single site $i$, it is easier to start by putting $\mathcal{O}$ under a unitary transformation (green triangle). Applying $U$ once, $\mathcal{O}$ only affects the unitary transformation and two adjacent controllers, and other part of the circuit at this layer can be efficiently contracted.  We obtain a two site operator which is purely composed of projection operator in the next layer.  Since the unitary transformation later on does not effect site $i$ anymore, we are ready to compute its expectation value, which is just a number entering the next layer. Putting the words above into equations, we have
\begin{equation}
\begin{aligned}
\tilde{\mathcal{O}}^{(1)}&=\bra{\rightarrow_i}u_i^{}\mathcal{O}u_i^{\dagger}\ket{\rightarrow}\\
                             &=\sum\limits_{s_{i-1},s_{i+1}}\mathcal{P}_{s_{i-1}}\bra{\rightarrow}u^{}_{s_{i-1}s_{i+1}}\mathcal{O}u^{\dagger}_{s_{i-1}s_{i+1}}\ket{\rightarrow}\mathcal{P}_{s_{i+1}}\\
                             &=\mathcal{P}_\alpha m^{\alpha\beta}\mathcal{P}_\beta
\end{aligned}
\end{equation}
Here $\mathcal{P}$ is the projection operator. The superscript $(1)$ stands for the operator after applying $U$ once. Fig.~\ref{fig:single_site} (a) is a graphic representation of the formula above. If we apply the transformation again, there are two possible cases illustrated in Fig.~\ref{fig:single_site} (b). Either way, we retain an operator with the same form but with $m$ replaced by a new $\tilde{m}$.

More explicitly, the first case:
\begin{equation}
\mathcal{O}^{(2)}_1=\mathcal{P^\alpha}m^{\alpha\gamma}\bra{\rightarrow}u^{\alpha\beta} \mathcal{P}^\gamma u^{\alpha\beta\dagger}\ket{\rightarrow}\mathcal{P^\beta}
\end{equation}
The second case:
\begin{equation}
\mathcal{O}^{(2)}_2=\mathcal{P^\alpha}m^{\gamma\beta}\bra{\rightarrow}u^{\alpha\beta} \mathcal{P}^\gamma u^{\alpha\beta\dagger}\ket{\rightarrow}\mathcal{P^\beta}
\end{equation}
After averaging both cases, we obtain:
\begin{equation}
\tilde{m}^{\alpha\beta}=\frac{1}{2}\bra{\rightarrow}u^{\alpha\beta} \mathcal{P}^\gamma u^{\alpha\beta\dagger}\ket{\rightarrow}(m^{\alpha\gamma}+m^{\gamma\beta})
\label{eq:flow_single_m}
\end{equation}
If we obtain $m^{(n)}$ in the end of the unitary-RG transformation, then the targeted expectation value is:
\begin{equation}
\braket{\mathcal{O}}=\frac{1}{4}\sum_{\alpha\beta} m^{(n),\alpha\beta}
\end{equation}
All the information about the operator we are coarse-graining is encoded in the initial value of $m$.
\begin{equation}
m^{(1),\alpha\beta}=\bra{\rightarrow}u^{}_{\alpha\beta}\mathcal{O}u^\dagger_{\alpha\beta}\ket{\rightarrow}
\end{equation}
Writing down each component of Eq. \ref{eq:flow_single_m}, we have:
\begin{equation}
\begin{aligned}
\tilde{m}^{11}&=\frac{1}{4}(2(1+\tanh(2\beta J))m^{11}+(1-\tanh{2\beta J})(m^{12}+m^{21}))\\
\tilde{m}^{12}&=\frac{1}{4}(m^{11}+2m^{12}+m^{22})\\
\tilde{m}^{21}&=\frac{1}{4}(m^{11}+2m^{21}+m^{22})\\
\tilde{m}^{22}&=\frac{1}{4}(2(1+\tanh(2\beta J))m^{22}+(1-\tanh{2\beta J})(m^{21}+m^{12}))
\end{aligned}
\end{equation}
With Eq. \ref{eq:t_flow}, this set of equations completely defines an iteration procedure from the initial operator to the final fully normalized operator. The eigenvalues and eigenvectors for a single iteration of the mapping are:
\begin{equation}
\begin{aligned}
E=&
\begin{pmatrix}
1 & \frac{1}{2}(1+\tanh(2\beta J)) &\frac{1}{2} &\frac{1}{2}\tanh{2\beta J}
\end{pmatrix}\\
V=&
\begin{pmatrix}
1 & -1 & 0 & 1  \\
1 & 0 & -1 & -\frac{1}{1-\tanh(2\beta J)} \\
1 & 0 &  1 & -\frac{1}{1-\tanh(2\beta J)} \\
1 & 1 &  0 & 1
\end{pmatrix}
\end{aligned}
\end{equation}

Now we discuss behavior of the Hamiltonian under this unitary-RG transformation. Expanding the exponential in Eq.~\ref{eq:1d_is_ham}, the Hamiltonian is a transverse field model with next neighboring interaction:
\begin{equation}
\begin{aligned}
H=&-\sum\limits_{i}X_i-\sinh(2\beta J)\sum\limits_{ \vev{i,j}}Z_iZ_j\\
     &+\sinh^2(\beta J)\sum\limits_{\vev{\vev{i,j}}}Z_iZ_j+N\cosh^2(\beta J)
\end{aligned}
\end{equation}
To obtain the new Hamiltonian, the strategy is to feed each term into our RG circuit while fixing the ancillary degree freedom into its ground state, namely all spin right. To make it clear, we assume that all ancillas are at even sites and the physical degrees of freedom are at odd sites.  Therefore for the $X_i$ term, there are two cases: even $i$ and odd $i$, the renormalized form of which are different. For the first case:
\begin{equation}
\begin{aligned}
X_{2m+1}&\rightarrow \mathcal{P}^\alpha_{2m} \bra{\rightarrow}u^{\alpha\beta}_{2m+1}X_{2m+1}u^{\alpha\beta\dagger}_{2m+1}\ket{\rightarrow}\mathcal{P}^\beta_{2m+2}\\
     &=\(\frac{1}{\cosh(2\beta J)}-1\)(\mathcal{P}^\uparrow_{2m}\mathcal{P}^\uparrow_{2m+2}+\mathcal{P}^\downarrow_{2m}\mathcal{P}^\downarrow_{2m+2})+1
\end{aligned}
\end{equation}
The second case is more involved:
\begin{equation}
\begin{aligned}
X_{2m}&\rightarrow\mathcal{P}_{2m-2}^{\alpha}\bra{\rightarrow}u^{\alpha\beta}_{2m-1}u^{\alpha\gamma\dagger}_{2m-1}\ket{\rightarrow}\mathcal{P}^{\beta}_{2m}X\mathcal{P}^{\gamma}_{2m}\\
     &\bra{\rightarrow}u^{\beta\delta}_{2m+1}u^{\gamma\delta\dagger}_{2m+1}\ket{\rightarrow}\mathcal{P}^{\delta}_{2m+2}\\
     &=\frac{\cosh^2(\beta J)}{\cosh(2\beta J)}X_{2m}
\end{aligned}
\end{equation}
The transformation of the nearest neighboring interaction also has two cases, but it turns out that the two cases are identical:
\begin{equation}
\begin{aligned}
Z_{2m-1}Z_{2m}&\rightarrow\mathcal{P}^\alpha_{2m-1}Z_{2m-1}\bra{\rightarrow}u^{\alpha\beta}_{2m}Z_{2m}u^{\alpha\beta\dagger}_{2m}\ket{\rightarrow}\mathcal{P}^\beta_{2m+1}\\
                          &=\tanh(2\beta J)(\mathcal{P}^{\uparrow}_{2m-1}\mathcal{P}^{\uparrow}_{2m+1}+\mathcal{P}^{\downarrow}_{2m-1}\mathcal{P}^{\downarrow}_{2m+1})\\
Z_{2m}Z_{2m+1}&=\tanh(2\beta J)(\mathcal{P}^{\uparrow}_{2m-1}\mathcal{P}^{\uparrow}_{2m+1}+\mathcal{P}^{\downarrow}_{2m-1}\mathcal{P}^{\downarrow}_{2m+1})\\
\end{aligned}
\end{equation}
The contribution of the two cases should add up and give an extra factor of 2.
Last we need study the transformation of the next neighboring interaction, which, as same as before has two cases. The first one:
\begin{equation}
\begin{aligned}
Z_{2m}Z_{2m+2}&\rightarrow \mathcal{P}^{\alpha}_{2m}Z_{2m}\mathcal{P}^{\beta}_{2m+2}Z_{2m+2}\\
                           &=2(\mathcal{P}^{\uparrow}_{2m}\mathcal{P}^{\uparrow}_{2m+2}+\mathcal{P}^{\downarrow}_{2m}\mathcal{P}^{\downarrow}_{2m+2})-1
\end{aligned}
\end{equation}
The second one:
\begin{equation}
\begin{aligned}
Z_{2m-1}Z_{2m+1}\rightarrow &\mathcal{P}^{\alpha}_{2m-2}\bra{\rightarrow}u^{\alpha\beta}_{2m-1}Z_{2m-1}u^{\alpha\beta\dagger}_{2m-1}\ket{\rightarrow}\\
                                                &\mathcal{P}^{\beta}_{2m}\bra{\rightarrow}u^{\beta\gamma}_{2m+1}Z_{2m+1}u^{\beta\gamma\dagger}_{2m+1}\ket{\rightarrow}\mathcal{P}^{\gamma}_{2m+2}\\
                                              =&\tanh^2(2\beta J)(\mathcal{P}^{\uparrow}_{2m-2}\mathcal{P}^{\uparrow}_{2m}\mathcal{P}^{\uparrow}_{2m+2}+\mathcal{P}^{\downarrow}_{2m-2}\mathcal{P}^{\downarrow}_{2m}\mathcal{P}^{\downarrow}_{2m+2})
\end{aligned}
\end{equation}
Although these two terms look like they involve interactions between three $Z$s, they actually cancel each other, which is a necessary consequence of the $\IZ_2$ symmetry. After carefully organizing all the terms above, and converting $\beta$ into the renormalized $\tilde{\beta}$, one can find that the resultant Hamiltonian has the exact same form as it in Eq. \ref{eq:1d_is_ham} with an overall constant $\frac{1}{2}e^{-2\tilde{\beta}J}(1+e^{-2\tilde{\beta} J}$).

\section{2d Ising square root state}
\label{sec:ising-sqrt}

Having studied in detail the RG circuit for the square root state of the 1d statistical Ising model, we now turn to a construction of the RG circuit for the square root state associated with the 2d statistical Ising model. To carry out Module 1 for this model we will use a specific implementation of the real-space RG due to Levin and Nave
\cite{PhysRevLett.99.120601}. This procedure is already enough to give interesting results, so we focus on it for simplicity, but our considerations are sufficiently modular that they can be carried out for various extensions and generalizations of the original scheme.\footnote{Indeed, many improvements have been made upon
the tensor renormalization group (TRG) procedure described in \cite{PhysRevLett.99.120601}.
A few particularly successful innovations  are:
The addition of an extra step which takes into account
the environment of the tensors, called
SRG \cite{PhysRevLett.103.160601}, is numerically dramatically more successful.
It is not trivial to generalize the TRG to higher dimensions.
Generalizations which accomplish this goal include
HOSVD \cite{PhysRevB.86.045139} and the work \cite{PhysRevB.87.085130}.
More recently, schemes were proposed \cite{2014arXiv1412.0732E, 2015arXiv150205385E} which are designed to remove additional types of correlations not addressed by TRG and to produce a better approximation to scale invariance. The latter work used a tensor network RG scheme on a 2d statistical model to produce a MERA for a 1d quantum model (the statistical model being interpreted as the Euclidean path integral of the quantum model); this is distinct from our work, e.g. the statistical model is {\it not} the Euclidean path integral of the quantum square root state model.
It may, however, be usefully combined with our work, as we mention below. Even more recently, a possible further improvement has appeared \cite{2015arXiv151204938Y}.
}

To set up the model, put qubits on the links of the honeycomb lattice, and label
a basis by $ \ket{s}$, eigenstates of Pauli operators $Z_i$ on the links, $ Z_i \ket{s} = s_i \ket{s}$.
Consider the following square root state:
$$
\ket{\psi_T} =
{1\over \sqrt{{\cal Z}}}  \sum_{\{ s \}} \ket{ s } \sqrt{ T_{s_1s_2s_3} T_{s_3 s_4 s_5 } ... }
$$
with
$$ {{\cal Z}} \equiv \sum_{\{ s \}} TT... $$
the associated classical partition function.
As explained in \cite{PhysRevLett.99.120601},
if we take $T_{+++} = 1, T_{--+} = T_{+--} = T_{+--} = e^{ - 2 \beta J} $
(and other components of the tensor, which would describe domain walls which end, zero)
this is the Ising model on the triangular lattice (up to a factor of two in $\cal Z$), where the two link configurations represent:
$+ \equiv$  ``no domain wall" and
$- \equiv$ ``yes domain wall".
To turn on a magnetic field in the Ising model
(necessary to compute for example the magnetization)
requires a complication of this scheme which we do not write out explicitly.

Consider the state associated with the Ising model on any graph
\be\label{eq:ising-sqrt} \ket{h} \equiv { 1\over \sqrt{\cal Z}}  \sum_{ \{ s \}}
e^{ + {\beta J \over 2 } \sum_{\vev{ij} } s_i s_j } \ket{  \{ s \} } .\ee

(Note that we have chosen the normalization $h(s) =  -J \sum_{\vev{ij} } s_i s_j $
so that $J>0$ gives a ferromagnetic classical ising model.)

Acting on qubits at the sites of any graph (not just 2d lattices), consider:
\be\label{eq:ising-h} {\bf H } \equiv \sum_i c_i(\beta) \( - X_i + e^{ -  \beta J  Z_i \sum_{ \vev{i|j} }Z_j } \) ~~.\ee
The notation $ \vev{i|j} $ means ``the set of neighbors $j$ of the fixed site $i$".
$c_i(\beta)$ are positive coupling constants
the choice of which is discussed in \S\ref{sec:normalization}.

\def\ZZ{Z}
\def\XX{X}
The state $\ket{h}$ in \eqref{eq:ising-sqrt} is the groundstate of $ \HH$.
First of all, it is an eigenvector with eigenvalue zero, $ {\bf H } \ket{h} = 0 . $
In a little more detail,
\bea X_i  \ket{h}
&=& {1\over \sqrt{ \CZ} }
\sum_{\{ \tilde s_j \equiv s_j, j \neq i, \tilde s_i \equiv - s_i  \} }
  e^{  {\beta\over 2 } \sum_{\vev{ij} }  s_i s_j }  \ket{  \{ \tilde s \} }
\cr &=& {1\over \sqrt{ \CZ} }
\sum_{ \{ \tilde s \} }
  e^{- {\beta } \sum_{\vev{i|j} }  \tilde s_i \tilde s_j }
  e^{ + {\beta\over 2 } \sum_{\vev{ij} }  \tilde s_i \tilde s_j }  \ket{  \{ \tilde s \} }
\cr &=&  e^{ - \beta J  \ZZ_i \sum_{ \vev{i|j} } \ZZ_j }  \ket{h} .
\eea

Secondly, $ {\bf H}$ is positive, so the zero eigenvector is the groundstate.
In the sum over sites in ${\bf H}$, each term $ H_i$
is an operator with eigenvalues greater than or equal to zero.
This is because $H_i$ is block diagonal in the $Z$ basis for the neighbors;
in the block where $ \sum_{\vev{i|j}} Z_j\equiv S$, it is
$$ H_i  = - X_i + e^{ - \beta  J S Z_i } $$
which has eigenvalues $ 0, 2 \cosh \beta J S $.
The eigenvalues of ${ \bf H}$ itself are therefore bounded below by zero.
(This is an application of the Perron-Frobenius theorem.)

The physics of this model is more interesting than the corresponding 1d model. Here there are two phases, a paramagnetic phase at small $\beta$ and a ferromagnetic phase at large $\beta$. These phases are separated by a quantum critical point describing a symmetry breaking transition which is however not the usual Wilson-Fisher fixed point (it is not even conformally invariant). Because the exact critical point of the 2d statistical Ising model is known
(on the honeycomb lattice, for example, it is $ (\beta J)_\star \simeq 0.658 $ (\eg~\cite{Creswick:1992rp}))
we know the exact location of the critical point in the square root state model. We know this must be a quantum critical point because
the Hamiltonian is local but correlation functions of local operators, for example, $Z_j$, become long-ranged at this point. This critical point is a non-trivial interacting fixed point which is multicritical, meaning it has more than one relevant symmetry-preserving perturbation.
We say this because we know that the ordinary $z=1$ Wilson-Fisher
fixed point also lies on the same phase boundary between paramagnetic 
and ferromagnetic phases.
It would be interesting to understand a field theory description of this fixed point.

%
%
%

%

\subsection{RG circuit}

The RG step has two parts.
The first part is a channel-duality rewiring move, and the second is the
coarse-graining step.
In fact, both steps will involve ancilla qubits.

Let $\CH_a$ denote the single-qubit hilbert space of $a$.
The first step should be made of local unitaries which act by
$$U_1 :   \CH_{abcd} \otimes \CH_e \otimes \CH_f   \to \CH_{abcd} \otimes \CH_e \otimes \CH_f$$
We require:
$$ U_1
\left| \eqnfig{.4in}{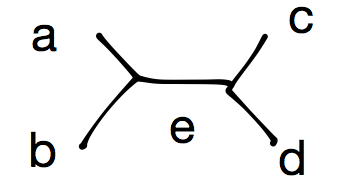}
~~~~~~\right\rangle\otimes \ket{0}_f
=
\sum_{f}
\left| \eqnfig{.4in}{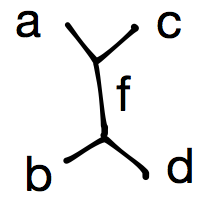}
\right\rangle\otimes \ket{0}_e
$$
or in more explicit notation,
$$ U_1 \sum_{abcde} \sqrt{ T_{abe} T_{ecd} } \ket{abcde} \otimes\ket{0}_f
 = \sum_{abcdf} \sqrt{ S_{acf} S_{fbd}  } \ket{abcd } \ket{0} \otimes \ket{f}_f~. $$
 Note that this rewiring move involves both adding and subtracting
ancillas.
To accomplish this, it suffices to take
 \be
 \label{eq:def-U1} U_1 \sum_e \sqrt{ T_{abe } T_{ecd}}\ket{abcde}\otimes\ket{0}_f
= \sum_f \ket{abcd0f} \sqrt{ { S_{acf} S_{fbd}  } }
 \ee
 Notice that we have not defined the action of $U_1$ on
 a general basis state.

As we demonstrate in \S\ref{sec:unitaries}, the classical RG relation
\be\label{eq:U1} \sum_{e}  T_{abe } T_{ecd} = \sum_f S_{acf } S_{fbd} \ee
is just what is needed to imply that $U_1$ is norm-preserving.


The second step is implemented by
$$ U_2
\left|
\parbox{.4in}{
\includegraphics[height=0.4in]{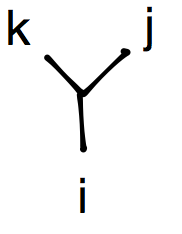}
}
\hskip-.1in
\right\rangle\otimes \ket{000}
=
\sum_{abc}
\left|
\parbox{.45in}{
\includegraphics[height=0.45in]{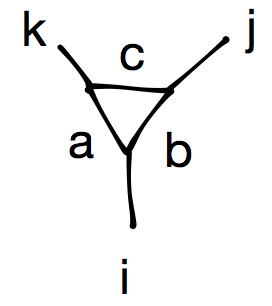}
}
\right\rangle
$$
The requirement is:
\be
\label{eq:def-U2}
 U_2 \sum_{ijk} \sqrt{ T_{ijk} T_{i..} T_{j..} T_{k..} } \ket{ijk}\otimes\ket{000}
= \sum_{ijkabc} \sqrt{ S_{iab} S_{jbc} S_{kac} T_{i..} T_{j..} T_{k..} } \ket{ijkabc} \ee
for all values of the un-named indices.
To accomplish this, it is sufficient simply to set
\be U_2 \ket{ijk000} =
\sum_{abc} \ket{ijkabc} \sqrt{  S_{iab} S_{jbc} S_{kac}  \over T_{ijk} } . \label{eq:U2guess} \ee
(The RHS is understood to be zero if any of the $S_{iab}$ vanish.)
Note that the relation
\be\label{eq:U2} \sum_{abc} S_{iab} S_{jbc} S_{kac}  = T_{ijk } \ee
implies that $U_2$ defined by this equation preserves the norm,
as shown in \S\ref{sec:unitaries}.

We note that the conditions \eqref{eq:def-U1} and \eqref{eq:def-U2}
do not completely specify $U_{1,2}$, since they do not determine the action
on excited states.  This is a useful freedom which merits further exploration.

\begin{figure}
\begin{center}
\includegraphics[width=.48\textwidth]{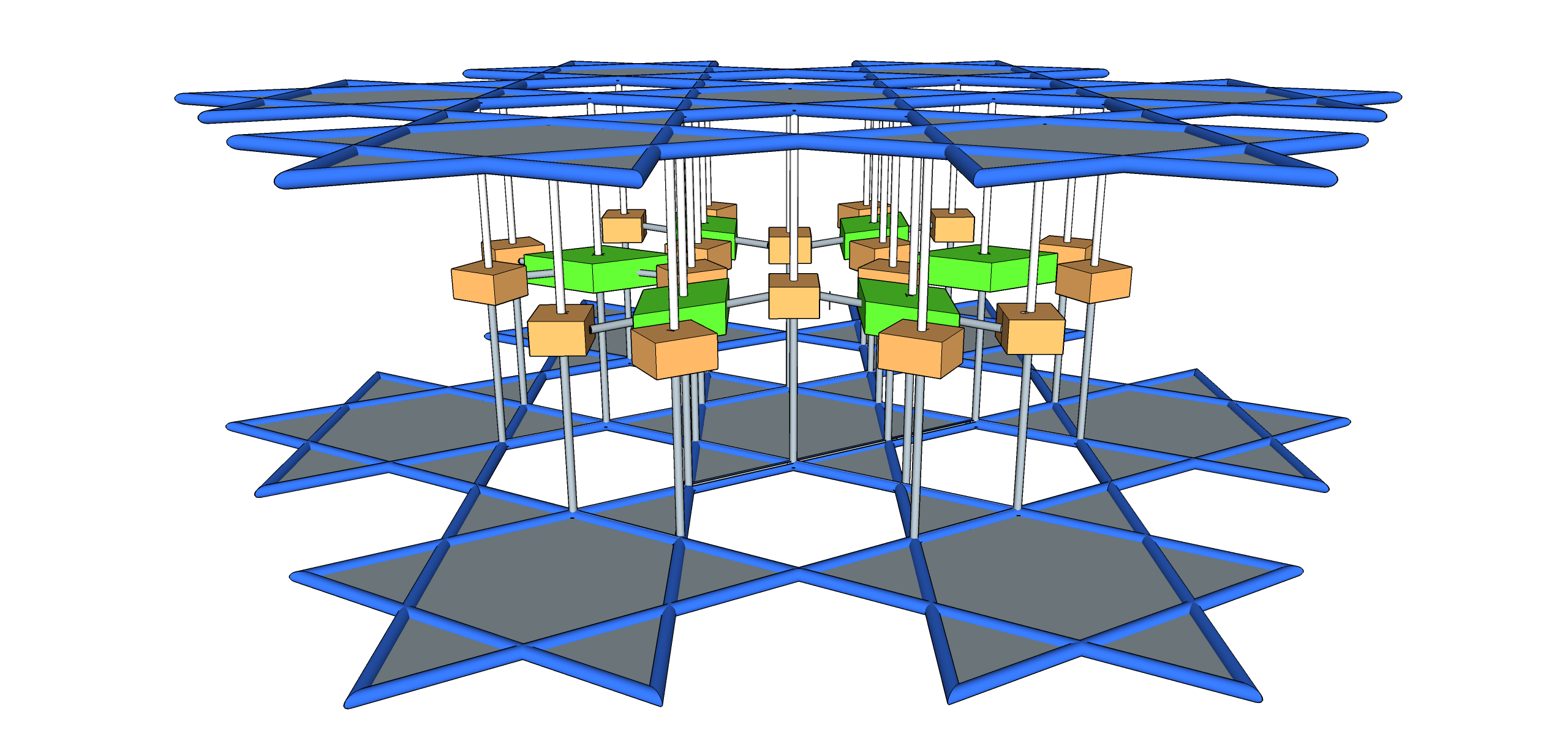}
\includegraphics[width=.51\textwidth]{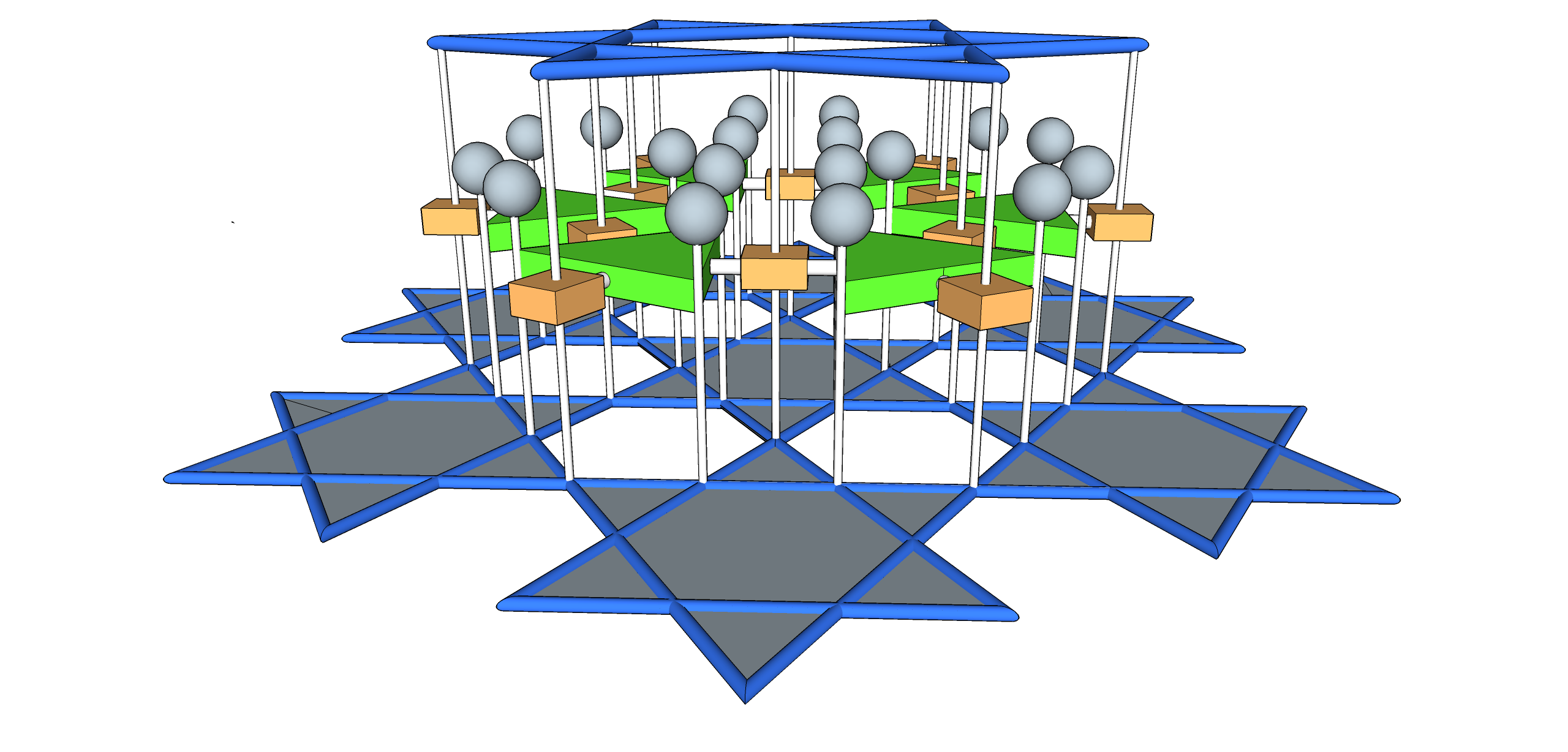}
\end{center}
\caption{A sketch of the ingredients of the RG circuit 
for a 2d quantum Ising square root state.
In the TRG representation of the triangular lattice Ising model, 
the tensors act on vector spaces associated with the links of the lattice;
we have found it convenient to 
draw directly the resulting link lattice,
which in this case is the Kagome lattice.
The rewiring step $U_1$ is at left and the disentangling step $U_2$ is at right.
Orange blocks are controllers, green blocks are unitary transformations that depend on the controllers that connected to them. White balls represent disentangled sites. This color choice is consistent with the figures
above for the one-dimensional case. 
}
\label{fig:kagome}
\end{figure}

The resulting unitary gates 
are depicted in Fig.~\ref{fig:kagome}.

\subsection{Truncation}

The procedure just outlined can {\it exactly} capture the critical point of the model if and only if an infinite bond dimension is used. However, we will show that a truncation to a rather modest bond dimension -- polynomial in system size -- is sufficient to guarantee high overlap with the true ground state in the thermodynamic limit. We need two crucial facts: (1) the scaling of entanglement in the quantum state described by the statistical model with boundary is logarithmic in subsystem size and (2) the particular sparse and conditional structure of the RG circuit produced above makes it easy to truncate the circuit while preserving unitarity.

Following
\cite{PhysRevLett.99.120601},
consider a large triangular region of the lattice, whose side lengths are $L$.
A sequence of coarse-graining maps on the wavefunction reduces the
product of tensors in this region to a single tensor with one index for each side of the triangle.
Fixing the values of the indices at the boundary of the region,
this product approaches (at large $L$) the groundstate wavefunction
of a 1d quantum system --
in the example on which we focus, it is the 1d transverse-field Ising model (TFIM).
Away from criticality, the $m$th eigenvalue of the reduced density matrix
of a subregion falls off like $ e^{ - c \log^2 m} $ for some constant $c$;
this holds as long as the subregion is much larger than the correlation length.
This falloff accounts for the favorable convergence of the TRG
away from the critical point \cite{PhysRevLett.99.120601}.

But even at criticality, the situation is not so dire.
The reduced density matrix for the state of the 1d quantum system
on each side of the triangle
has an eigenvalue distribution which is well-peaked about $ \log(\lambda) = -S$,
where $S$ is its von Neumann entropy
\cite{2013arXiv1304.6402S,Czech:2014tva}.
Therefore, there exists a number $k$ of order one such that truncating the infinite bond dimension to $ e^{k S}$ states incurs only a small error of order $ e^{-S}$.
For the groundstate of the critical TFIM, a 1d CFT with central charge $c=1/2$, the entanglement entropy of an interval of length $L$ behaves as $S(L) = {c \over 3} \log L $ \cite{Holzhey:1994we}. Thus with a truncated bond dimension of size $e^{k S}$, that is polynomial in $L$, the error in our approximation to the groundstate of the large triangle
goes like $ e^{ -  S } = L^{- c/3 } $.

It is also important that the truncated circuit with bond dimension $e^{kS}$ is still composed of unitary operators.
The crucial conditions are \eqref{eq:U1} and \eqref{eq:U2},
which must be satisfied with the summations
running over the appropriate finite bond dimension.

The conditions \eqref{eq:U1} and \eqref{eq:U2} can be solved numerically 
with arbitrary $\beta$, using various bond dimensions, as in 
\cite{PhysRevLett.99.120601}. 
It will be interesting to use the resulting data  
to learn more about scaling dimensions of operators
at the critical point.

\subsection{Topologically ordered phase}

In fact, the Ising models we have been considering, when placed on the right kind of lattice, can describe even more interesting phases. 
This will allow us to make contact with previous literature on exact RG circuits \cite{2009PhRvB..79h5118G, 2008PhRvL.100g0404A}.

On any bipartite lattice
a sublattice rotation
$Z_i \to (-1)^i Z_i$
relates $J>0$ to $J<0$
for the $e^{ - \beta J Z_i \sum Z_j }$ term of \eqref{eq:ising-sqrt},
just as it relates ferromagnetic and antiferromagnetic (AFM) statistical Ising models.
But on a non-bipartite lattice there is something different at $J<0$.
In the state associated with a classical frustrated magnet,
there are many terms in the superposition with the same weight.
This is a symptom of topological order.
In particular,
there is a map from the triangular lattice AFM to the honeycomb lattice dimer model:
the domain walls on the honeycomb lattice form closed loops which should be regarded as {\it differences} of dimer configurations.
(For a summary of this mapping, see appendix A of
\cite{2002PhRvB..65b4504M}
and \cite{PhysRevB.64.144416}.)

Consider quantum spins on the triangular lattice.
States which makes the antiferromagnetic Heisenberg interaction locally happiest
have one link of each triangle in a singlet.
Such states can be mapped to dimer coverings (every site covered by exactly one dimer)
of the dual (honeycomb) lattice just by covering the links which intersect the singlets.
The uniform superposition of these states is closely related to the state we get in the limit
$ \beta \to \infty, \beta J < 0 $.
The only difference is that instead of singlets, we have $ \ket{ \up \down } + \ket{\down \up}$ -- all positive coefficients --
on the links which disagree.
This difference is of the form described in \S\ref{sec:multiple-roots},
taking advantage of the ambiguity in the phase of the square root.
So this limit gives exactly the Rokhsar-Kivelson state \cite{Rokhsar:1988zz}.

Since $\beta = 0 $ is a paramagnet there must be another phase transition in between at negative $J$.

In the limit $ \beta \to \infty, \beta J < 0 $, the construction above is exact, with finite bond dimension.
In particular, the tensors simplify dramatically:
with the labelling where the index $ i$ counts the number of domain walls on the associated link 
($i=0$ or $1$), we have 
$$ T_{ijk} = \delta_{i+j+k} $$
where the argument of the Kronecker delta is to be understood modulo two:
it merely enforces that the domain walls are closed loops.

The resulting circuit is self-dual under channel duality: 
$$ \sum_{k=0}^1 T_{ijk} T_{klm} = \sum_{n=0}^1 T_{ikn} T_{njm} $$
-- that is $ S=T$.
In this limit, the rewiring move 
$$ U_1
\left| \eqnfig{.4in}{fig-channel-t.png}
~~~~~~\right\rangle\otimes \ket{0}_f
=
\sum_{f}
\left| \eqnfig{.4in}{fig-channel-s.png}
\right\rangle\otimes \ket{0}_e
$$
is accomplished by 
$ U_1 = CX_{ae}CX_{be} CX_{af} CX_{cf} $, 
where the control-X gate is $ CX_{12} \equiv \sum_{s_1} \ket{s_1} \bra{s_1} \otimes X_2^{s_1} $.
This is a result of \cite{2009PhRvB..79h5118G}.
Similarly, the decimation move 
$$ U_2
\left|
\parbox{.4in}{
\includegraphics[height=0.4in]{fig-star-triangle2.png}
}
\hskip-.1in
\right\rangle\otimes \ket{000}
=
\sum_{abc}
\left|
\parbox{.45in}{
\includegraphics[height=0.45in]{fig-star-triangle1.png}
}
\right\rangle
$$
is accomplished by 
$$
U_2 = CX_{kc} CX_{jc} CX_{ib} CX_{ab} CX_{ia} CX_{ka} ~.
$$
These more-specific formulae our consistent with the demands we put on our circuit.

\section{Discussion}
\label{sec:other}

In this paper we have provided examples of
quantum critical groundstates in various dimensions which
satisfy an area law and which have high-fidelity tensor network
representations with favorable (polynomial in system size) bond dimensions.
We anticipate that it is possible
to go beyond this result
to system-size independent bond dimension
using the new technology introduced in \cite{2014arXiv1412.0732E}.

In appendix \ref{sec:other-models}, we 
formulate
square root states for classical models 
with long-range interactions.
In the rest of this concluding section, we briefly
discuss other directions in
which one might apply the technology developed here.

\subsubsubsection{Quantum Lifshitz theories and generalizations}

It is not necessary that the configuration space 
of the classical model be discrete.  For example, it may be a continuum field theory.
We recall the structure of the``Lifshitz theories" described in
\cite{Ardonne:2003wa}
(and more recently studied in \cite{Horava:2009uw})
where the stat mech model in question is a Gaussian free field.
In particular, there we have states labelled by a configuration of a
scalar field $\phi(x)$.
(The continuum is not so crucial, but the notation is nicer.)
$$ \ket{h}
= \int [D\phi] \ket{\phi}
e^{ - \half \int d^d x ( \nabla \phi)^2 }
\equiv \int [D\phi] \ket{\phi} \Psi[\phi]
.$$
Since
$$ - { \delta \over \delta \phi(x) } \Psi[\phi]
= +  { \delta \over \delta \phi(x) }\int d^d x ( \nabla \phi)^2 \Psi[\phi]
= - \nabla^2 \phi(x)  \Psi[\phi] $$
the wavefunction satisfies
$$\(-  \(  { \delta \over \delta \phi(x) } \)^2 + \( \nabla^2 \phi \)^2 \) \Psi[\phi]  = 0 .$$
Since the operator
$$ \HH \equiv  \int d^dx\(    \pi(x)^2 +  \( \nabla^2 \phi (x)\)^2  \)$$
(here $ \pi(x) $ is the canonical field momentum, $ [ \phi(x), \pi(y)] = i \delta^d(x-y) $)
is positive,
the state $\ket{h}$ with eigenvalue zero is its groundstate.

More generally,
it's not so important that the classical $h$ be quadratic.
We could replace $\int (\nabla \phi)^2 $ with
any real local functional
$S[\phi]$
and the state
$$ \ket{S}
= \int [D\phi] \ket{\phi}
e^{ - S[\phi] }
$$
is the
groundstate  of
$$ \HH_S \equiv   \int d^d x \( \pi(x)^2 +  \(  { \delta S \over \delta \phi(x) }\)^2   +  { \delta^2 S \over \delta \phi(x)^2 }\).  $$

\subsubsubsection{Multiple roots}
\label{sec:multiple-roots}

Our construction has numerous extensions.
For example: as always, there is more than one square root.
Since $\langle s | s' \rangle = \delta_{ss'}$, we can multiply each basis state $\ket{s}$ by an
$s$-dependent phase without losing the defining property
that correlators of $Z$-basis operators in the state are given by the classical model.

So a much larger class of square root states is of the form
$$ \ket{h, \phi}  \equiv {1\over \sqrt Z } \sum_s e^{ - \half \beta h(s) } e^{ i \phi(s) } \ket{s} $$
where $\phi$ is any real function on the stat mech configuration space.

Positivity of the wavefunction at $\phi =0$ is useful
for application of the Frobenius theorem,
and in general this is lost for $\phi \neq 0$.
These states can certainly be orthogonal to $\ket{h, \phi=0}$.

Correlation functions of $Z$s are independent of $\phi$, 
because the absolute value removes this phase from each term of the sum.
However, correlations of off-diagonal operators involving $X$s will depend on $\phi$.

This suggests a further generalization: we may consider square root states of 
partition functions which are sums of complex weights.
Such sums arise for example in the euclidean path integral formulation
of quantum systems with nontrivial Berry phases.

\subsubsubsection{Dynamics}

While most of this paper has focussed on groundstate properties,
of course dynamics are interesting too.
The frustration-free construction
we have employed
means we don't learn that much about dynamics from the groundstate.
In particular, there are many local Hamiltonians with this same groundstate,
but different spectra of excited states away from criticality.
(For every such choice, the gap must close at the critical point.)

However, we can say something about the dynamics for some natural
choice of the Hamiltonian,
as we describe in appendix \ref{sec:bound-on-z}.
Specifically, it is possible to bound the dynamical critical exponent from below.
We leave it for the future to use the RG circuit constructed above
to determine its precise value.



\vskip.2in
{\bf Acknowledgements.}
Thanks to Dan Arovas and Ning Bao for discussions
and Diptarka Das for comments on the manuscript.
This work was supported in part by
funds provided by the U.S. Department of Energy
(D.O.E.) under cooperative research agreement 
DE-SC0009919.
BS is supported by funds from the Simons Foundation and Stanford Institute for Theoretical Physics.
SX is supported by NSF DMR-1410375 and AFOSR FA9550-14-1-0168

\appendix
\renewcommand{\theequation}{\Alph{section}.\arabic{equation}}

\section{Long range interactions in the classical model}
\label{sec:other-models}


Let us consider somewhat non-local classical hamiltonians.
A motivation for attempting this is that the ground state of say a relativistic scalar field is positive definite and can be thought of as the square root of some statistical weight, but that weight will have power law decaying interactions if the field is massless.

Let $ h = \sum_{ij} J_{ij} s_i s_j $ (with e.g. $ J_{ij} \sim {1\over |d(i,j)|^\alpha} $ ),
so the classical partition sum is
$$ \CZ = \sum_s e^{ - \beta h(s) }
\propto \sum_s \int \prod_j dX_j e^{ -  \beta (  \ii \sum_i s_i X_i  + {1\over 4} \sum_{ij} X_i J^{-1}_{ij} X_j ) } $$
where $J^{-1}$ is the matrix inverse of $J$.

The associated quantum state is:
$$\ket{h}
= { 1\over \sqrt{\CZ}} \sum_s e^{ - \beta h(s)/2} \ket{s}
\propto \int [DX] e^{ {\beta\over 4 } X_i J^{-1}_{ij} X_j}
\otimes_k ( e^{ -  \ii \beta X_k } \ket{\up}_{k} + e^{ \ii \beta X_k } \ket{\down}_k )  .$$

The introduction of the auxiliary field $X$ gives a tensor product state:
$$\vev{s|h} \propto \int [DX] e^{-  {\beta\over 4 } X_i J^{-1}_{ij} X_j}
 e^{ -\ii  s_k \beta X_k } ,$$
\ie~it is a sum of product states where the local spin direction in each term is determined
by the local auxiliary field.
The auxiliary field acts like a local (imaginary) magnetic field.

Now any RG we know how to do on the $X$ path integral
tells us how to coarse-grain the state.

 \subsubsubsection{Quantum Laughlin plasma analogy}

Another example which fits in this framework is the Laughlin wavefunction for
incompressible abelian fractional quantum Hall states \cite{PhysRevLett.50.1395}.
The stat mech model for that case is the plasma of the  ``plasma analogy", \ie~a 2d classical gas of particles with logarithmic forces.
This example
seems different from the spin examples
because the wavefunction in question is in a state of definite particle number, in position space.
Thinking of it this way gives a derivation of the associated Moore-Read CFT.

The norm of the Laughlin wavefunction at filling $ \nu = 1/m$ is
$$
\langle z_1..z_N| \text{laughlin}_m \rangle =  \prod_{i<j} |z_i - z_j|^{2m} e^{ - \sum_{i} |z_i|^2/ 2 l^2 } \equiv e^{ - h(z) }
$$
with
$h(z) = \sum_{i=1}^N |z_i|^2/4 l^2 - m \sum_{i<j } \log |z_{ij}|^2  $.  $l$ is the magnetic length.

Usually one just thinks about the plasma analogy for the norm.
But let's write the wavefunction itself using a lagrange multiplier
to make the interaction in $h(z)$ local (in the $z$ space):

\bea | \text{laughlin}_m \rangle &=& \int d^N z \ket{ z_1 .. z_N } \prod_{i<j} |z_{ij}|^m e^{ - \sum_{i} |z_i|^2/ 4 l^2 }
\cr
&=&
\nonumber
\int d^Nz e^{ - \sum_{i} |z_i|^2/ 2 l^2 }  \ket{ z_1 .. z_N }
\int [D\phi(z)] e^{ - \int d^2 z  \( {1\over 4 \pi}  \partial_z \phi(z, \bar z) \bar \partial_z \phi(z,\bar z)
+ \phi(z,\bar z) \rho(z, \bar z ) \)}  \eea
where the source is $ \rho(z, \bar z)  = m \sum_{i=1}^N \delta^2(z- z_i ) $.
This is the $c=1$ theory whose correlators (by construction now!) give the wavefunction.
that is:
$$ \langle z_1..z_N| \text{laughlin}_m \rangle
= e^{ - \sum_{i} |z_i|^2/ 2 l^2 }
 \int [D\phi(z)] e^{ - \int d^2 z  \( {1\over 4 \pi}  \partial_z \phi(z, \bar z) \bar \partial_z \phi(z,\bar z)
+ \phi(z,\bar z) \rho(z, \bar z )\) } $$
(Actually, we've lied a little bit above: a single copy the wavefunction itself is only the chiral piece
of a free boson, 
whose path integral representation is a little problematic -- 
it requires an extension of the configuration to an extra dimension and 
the use of the Chern-Simons action.)

Notice that in this case, the associated stat mech model is an RG fixed point,
despite the fact that the state in question is gapped --
like known scale-invariant MERAs for non-chiral topologically-ordered gapped state.

\section{Normalization of the Ising square root Hamiltonian and the limit $T\to 0$}
\label{sec:normalization}


The constants $c_i(\beta)$
in the normalization of the Hamiltonian \eqref{eq:ising-h}
do not affect the statement that $ \ket{h}$ is a groundstate.
But they can be chosen to make the $\beta \to \infty$ zero-temperature limit
more uniform.
In particular,
notice that
$$ e^{ -  \half \beta \ZZ_1 \ZZ_2 } c_0(\beta) \ket{ \up\down} = c_0(\beta) e^{+ \half \beta} \ket{\up\down},
~~~~
e^{ -  \half \beta \ZZ_1 \ZZ_2 } c_0(\beta) \ket{ \up\up} = c_0(\beta) e^{-\half \beta} \ket{\up\up}
$$
so if we choose $c_0(\beta) = e^{ - \beta}$ the first expression stays finite as $\beta \to \infty$:
$$ e^{- \half \beta \ZZ_1 \ZZ_2 } e^{- \beta} = e^{ -  \beta \PP_{0}(\ZZ_1 \ZZ_2) } $$
where $\PP_{0}(Z)$ is the projector onto $ Z = 1$.
Since $\PP_0^2 = \PP_0$, we have
$$ e^{ - \beta \PP_0 }  = (1 - \PP_0 ) + e^{ - \beta}  \PP_0
= \PP_1 + e^{ - \beta} \PP_0  $$
($\PP_1(Z) $ projects onto $ Z= - 1$).

So we are led to take
$$ c_i(\beta) = e^{ - \half \beta J n_i } $$
where $n_i$ is the degree of the site $i$ (\ie~the number of neighbors),
and the hamiltonian can be written as:
\bea {\bf H }  &=&   \sum_i   \( - e^{ - \half n_i \beta J }  \XX_i +
\prod_{\vev{i|j}} e^{  - \beta J \PP_{0}(\ZZ_i \ZZ_j ) }  \)
\cr
&= &
 \sum_i   \( - e^{ - \half n_i \beta J }  \XX_i +
\prod_{\vev{i|j}} \(\PP_1(\ZZ_i \ZZ_j)  + e^{ - \beta} \PP_0 ( \ZZ_i \ZZ_j)
 \)   \) ~
\eea
Notice that in the $\beta \to \infty$ limit,
the paramagnetic term goes away.
Further, the remaining term becomes just
$$ \lim_{\beta \to \infty} \HH
= \sum_i \prod_{\vev{i|j}} \PP_1(\ZZ_i \ZZ_j)  ~.
$$
This exacts a penalty for any disagreement between neighboring spins,
and is zero on states where all the spins agree.
This is consistent with the fact that the state
$ \ket {h} $ reduces to
$$\lim_{\beta \to + \infty } \ket{h} =  {1\over \sqrt{2 } }\(\ket{ \up \up\up\up...} + \ket{ \down\down\down \down...} \)
$$
in this limit.

\section{Bounding the dynamical exponent of the critical 2d Ising square root state}
\label{sec:bound-on-z}

Here is a variational bound on the dynamical critical exponent
of the 2d Ising square-root quantum critical point.
Briefly, it can be described as using the single-mode approximation as a variational state.

%
%
%

Consider the ansatz
$$ \ket{\phi} = \sum_i Z_i \ket{gs} \equiv M \ket{gs}  . $$
This state has the opposite eigenvalue of $ \prod_i X_i $ from the groundstate.
The energy expectation in this state provides an
upper bound on the energy of the first excited state.
This follows if we know that the first excited state is in the other symmetry sector.
(Exact diagonalization on small systems indicates this to be true but a proof
has not materialized.)

Its norm is
$$ \vev{ \phi| \phi} = Z \vev{M^2 }_\text{ising} \sim L^{4-\eta} Z $$
where the last relation holds at the critical point, and $\eta = 1/4$ is the
twice the order parameter critical exponent.

So the lowest energy in the wrong-symmetry sector must be below
$$ \frac{ \bra{\phi} H \ket{ \phi }} { \vev{\phi | \phi} }
\sim { L^2 \over L^{4-\eta} Z }  E_i $$
where $E_i  = \bra{\phi} H_i \ket{\phi}$
is the expectation for a single term in $H$.
The latter can be written as
$$ E_i = \sum_s \vev{ \phi | s } \sum_{s'} \bra{s } H_i \ket{s'} \vev{ s'  |\phi} 
$$
Using $  \vev{ \phi | s }  = M_s w_s^{1/2} $ where $w_s = e^{ - \beta \sum_{\vev{ij} } s_i s_j } $
and $M_s = \sum_i s_i $, this is
$$ E_i =  \sum_s M_s w_s^{1/2} \( e^{ \beta s_i \sum_{\vev{i|j} }s_j } M_s w_s^{1/2}
- M_{s'} w_{s'}^{1/2} \)  $$
where $s$ and $s'$ differ by flipping $s_i$, so that
(as in the construction of $H$)
$$ w_{s'}^{1/2}  = w_s^{1/2} e^{ \beta s_i \sum_{\vev{i|j} }s_j } .$$
So
$$ E_i =  \sum_s M_s w_s e^{ \beta s_i \sum_{\vev{i|j} }s_j } \(  M_s
- M_{s'}  \)  $$
Now note that
$$ w_s e^{ \beta s_i \sum_{\vev{i|j}} s_j }  = w_{s \setminus s_i }$$
where the RHS is the weight without the links containing the site $i$.
Also:
$$ M_s = \sum_{j\neq i} \(s_j + s_i \), ~~ M_s - M_{s'} = 2 s_i .
 $$
So
$$ E_i = \sum_{ \{ s \} \setminus s_i }  w_{ s\setminus s_i }
\sum_{s_i = \pm } \(  \sum_{j\neq i}s_j  + s_i\)  2 s_i
= 2 Z( \setminus i ) $$
where the RHS is the partition function of the ising model with the site $i$ removed.

This quantity
$$ Z( \setminus i )  = Z \vev{ e^{ \beta s_i \sum_{\vev{i|j} } s_j } } $$
is bounded (on a lattice with coordination number 4 )
by
$$ Z e^{ - 4 \beta}  < Z Z( \setminus i ) < Z  e^{ + 4 \beta}  .$$
This means that at large $L$ it must be a positive constant  times $Z$.

Therefore:
the scaling of the excited state energy at the critical point is bounded above by
$$ E_1 = { c' \over L^z }  > { c \over L^{2 - \eta} } , $$
and hence the dynamical exponent is bounded below by
$$ z  \geq 2 - \eta = 1.75 .$$

\section{Unitarity check}\label{sec:unitaries}

Unitary operators are in particular inner-product-preserving.  Here we check explicitly
that this property follows by construction for our unitaries
made from the Levin-Nave RG tensors.
Beginning from the ansatz (\ref{eq:U2guess}) the goal is to check
\be
\langle 000 i j k| U_2^\dagger U_2 |i' j' k' 000\rangle \stackrel{?}{=} \delta_{i, i'} \delta_{j, j'} \delta_{k,k'}.
\ee

From the definition \eqref{eq:U2guess}, we have
$$ \langle  i j k000| U_2^\dagger
= \sum_{ijk abc} \bra{ijkabc}  \sqrt{\frac{ S_{abi}  S_{bcj} S_{cak}  }{T_{ijk} }}$$
(Note that we are using a convention where the arguments of the bra are in the same order as in the ket,
and for simplicity we are assuming $S,T$ are real.)
Therefore the inner product
$$ \vev{ ijk000 | i'j'k' 000} =  \delta_{i, i'} \delta_{j, j'} \delta_{k,k'}.$$
maps to
\bea
\langle  i j k000| U_2^\dagger U_2 |i' j' k' 000\rangle
&=& \sum_{abc} \sum_{a'b'c'}
\underbrace{\langle   i j k abc | i' j' k' a'b'c'\rangle}_{
= \delta_{aa'}\delta_{bb'} \delta_{cc'} \delta_{ii'} \delta_{jj'} \delta_{kk'} }    \sqrt{\frac{S_{abi}S_{a'b'i'} S_{bcj}S_{b'c'j'} S_{cak} S_{c'a'k'}}{T_{ijk} T_{i'j'k'}}}
\cr  ~\cr  \cr
&=&  \delta_{ii'} \delta_{jj'} \delta_{kk'}
\sum_{abc}
\frac{S_{abi} S_{bcj} S_{cak} }{T_{ijk} }
\cr
&=& \delta_{i, i'} \delta_{j, j'} \delta_{k,k'} = \vev{ijk000|i'j'k'000}.
\eea

For $U_1$ we have
 \be  \sum_e   \bra{abcde}\otimes\bra{0}_f \sqrt{ T_{abe } T_{ecd}} U_1^\dagger
= \sum_f \bra{abcd0_ef} \sqrt{ { S_{acf} S_{fbd}  } }
 \ee
 So
 \bea
\sum_{e,e'} \sqrt{ T_{a'b'e' } T_{e'c'd'} T_{abe } T_{ecd} }\bra{a'b'c'd'e'0_f} &U_1^\dagger &U_1 \ket{abcde0_f}
\cr
&=& \sum_{f,f'}  \underbrace{\vev{a'b'c'd'0_ef' | abcd0_ef}}_{\delta_{abcdf}^{a'b'c'd'f'}}
\sqrt{  S_{a'c'f'} S_{f'b'd'} S_{acf} S_{fbd} }
\cr
&=&
 \delta_{abcd}^{a'b'c'd'} \underbrace{\sum_f S_{acf} S_{fbd}}_{  = \sum_{e}  T_{abe }  T_{ecd}}
\cr &=&
\sum_{e,e'} \sqrt{ T_{a'b'e' } T_{e'c'd'} T_{abe } T_{ecd} }\vev{a'b'c'd'e'0_f|abcde0_f}
\nonumber
\eea


\bibliographystyle{ucsd}
\bibliography{collection}
\end{document}

%% file: jm-tex-macros-public.tex
\usepackage{amssymb}
\usepackage{amsmath,bm}
\usepackage{amssymb}
\usepackage{graphicx}
\usepackage{amsfonts}         
\usepackage{fancybox}

\usepackage{enumitem}

\usepackage{slashed}

\usepackage[font=small,labelfont=bf]{caption}
\DeclareCaptionFont{tiny}{\tiny}
\usepackage{yfonts} 
\usepackage{epstopdf}
\usepackage[usenames,dvipsnames]{xcolor}
\usepackage[pdftex, bookmarks={false}, pdfauthor={John McGreevy}, pdftitle={This is a secret message!}]{hyperref}
\hypersetup{colorlinks=true, linkcolor=BrickRed, citecolor=Violet, filecolor=OliveGreen, urlcolor=RoyalBlue, filebordercolor={.8 .8 1}, urlbordercolor={.8 .8 0}}
\usepackage{soul}
\setstcolor{Red}

\usepackage{wrapfig}

\definecolor{darkgreen}{rgb}{0,0.4,0}
\definecolor{darkred}{rgb}{0.4,0,0}
\definecolor{darkblue}{rgb}{0,0,0.4}
\definecolor{lightblue}{rgb}{.6,.6,0.9}

\definecolor{uglybrown}{rgb}{0.8,  0.7,  0.5}

\definecolor{palatinatepurple}{rgb}{0.41, 0.16, 0.38}
\definecolor{celebrationcolor}{rgb}{0.75,  0.0,  0.9}

 \usepackage{mdframed}

\usepackage{framed}
\definecolor{shadecolor}{rgb}{0.90,0.90,0.90}

\usepackage{wasysym}

\def\subsubsubsection#1{{\bf #1}}

\numberwithin{equation}{section}

\renewcommand{\theequation}{\arabic{section}.\arabic{equation}}

\def\nd{{ \vphantom{\dagger}}}

\input epsf
\newcommand{\vev}[1]{\langle #1 \rangle}

\newlength{\extraspace}
\newlength{\extraspaces}
\def\be{\begin{equation}}
\def\ee{\end{equation}}

\newcommand{\bea}{\begin{eqnarray}}
\newcommand{\eea}{\end{eqnarray}}

\def\half{{1\over 2}}

\def\tr{{\rm tr}}

\def\bra#1{\left\langle#1\right|}
\def\ket#1{\left|#1\right\rangle}

\def\vev#1{\left\langle{#1}\right\rangle}

\def\CH{{\cal H}}

\def\CO{{\cal O}}

\def\CZ{{\cal Z}}

\def\II{\relax{I\kern-.10em I}}

\def\IZ{\mathbb{Z}}
\def\IB{\relax{\rm I\kern-.18em B}}

\def\ID{\relax{\rm I\kern-.18em D}}
\def\IE{\relax{\rm I\kern-.18em E}}
\def\IF{\relax{\rm I\kern-.18em F}}
\def\IG{\relax\hbox{$\inbar\kern-.3em{\rm G}$}}
\def\IGa{\relax\hbox{${\rm I}\kern-.18em\Gamma$}}
\def\IH{\relax{\rm I\kern-.18em H}}
\def\II{\relax{\rm I\kern-.18em I}}
\def\IK{\relax{\rm I\kern-.18em K}}

\def\inbar{\,\vrule height1.5ex width.4pt depth0pt}

\def\lp10{\ell_p^{10}}
\def\lp11{\ell_p^{11}}
\def\R11{R_{11}}

\def\frac#1#2{{#1 \over #2}}

\def\up{\uparrow}
\def\down{\downarrow}

\def\Ione{\hbox{$1\hskip -1.2pt\vrule depth 0pt height 1.53ex width 0.7pt
                  \vrule depth 0pt height 0.3pt width 0.12em$}}

\newdimen\tableauside\tableauside=1.0ex
\newdimen\tableaurule\tableaurule=0.4pt
\newdimen\tableaustep
\def\phantomhrule#1{\hbox{\vbox to0pt{\hrule height\tableaurule width#1\vss}}}
\def\phantomvrule#1{\vbox{\hbox to0pt{\vrule width\tableaurule height#1\hss}}}
\def\sqr{\vbox{%
  \phantomhrule\tableaustep
  \hbox{\phantomvrule\tableaustep\kern\tableaustep\phantomvrule\tableaustep}%
  \hbox{\vbox{\phantomhrule\tableauside}\kern-\tableaurule}}}
\def\squares#1{\hbox{\count0=#1\noindent\loop\sqr
  \advance\count0 by-1 \ifnum\count0>0\repeat}}
\def\tableau#1{\vcenter{\offinterlineskip
  \tableaustep=\tableauside\advance\tableaustep by-\tableaurule
  \kern\normallineskip\hbox
    {\kern\normallineskip\vbox
      {\gettableau#1 0 }%
     \kern\normallineskip\kern\tableaurule}%
  \kern\normallineskip\kern\tableaurule}}
\def\gettableau#1 {\ifnum#1=0\let\next=\null\else
  \squares{#1}\let\next=\gettableau\fi\next}

 \def\eqnn#1{\xdef #1{(\secsym\the\meqno)}\writedef{#1\leftbracket#1}%
 \global\advance\meqno by1\wrlabeL#1}
 \def\eqna#1{\xdef #1##1{\hbox{$(\secsym\the\meqno##1)$}}
 \writedef{#1\numbersign1\leftbracket#1{\numbersign1}}%
 \global\advance\meqno by1\wrlabeL{#1$\{\}$}}
 \def\eqn#1#2{\xdef #1{(\secsym\the\meqno)}\writedef{#1\leftbracket#1}%
 \global\advance\meqno by1$$#2\eqno#1\eqlabeL#1$$}

\global\newcount\itemno \global\itemno=0

\def\itemaut#1{\global\advance\itemno by1\noindent\item{\the\itemno.}#1}

\def\({\left(}
\def\){\right)}

\def\ii{{\bf i}}

\def\HH{{\bf H}}

\def\PP{{\bf P}}

\def\UU{{\bf U}}

 \def\XX{{\bf X}}
 \def\ZZ{{\bf Z}}

\def\lsim{\mathrel{\mathstrut\smash{\ooalign{\raise2.5pt\hbox{$<$}\cr\lower2.5pt\hbox{$\sim$}}}}}
\def\gsim{\mathrel{\mathstrut\smash{\ooalign{\raise2.5pt\hbox{$>$}\cr\lower2.5pt\hbox{$\sim$}}}}}

\def\overleftrightarrow#1{\vbox{\ialign{##\crcr
     $\leftrightarrow$\crcr\noalign{\kern-0pt\nointerlineskip}
     $\hfil\displaystyle{#1}\hfil$\crcr}}}
     
     \def\overleftarrow#1{\vbox{\ialign{##\crcr
     $\leftarrow$\crcr\noalign{\kern-0pt\nointerlineskip}
     $\hfil\displaystyle{#1}\hfil$\crcr}}}

\def\eg{{\it e.g.}}
\def\ie{{\it i.e.}}

\def\eqnfig#1#2{\parbox{#1}{
\includegraphics[height=#1]{#2}
}}

\usepackage[yyyymmdd,hhmmss]{datetime}
\newif{\ifeq}           
\eqtrue                 
\newcounter{lecturecounter}

%% file: no-revtex.tex
\renewcommand{\title}[1]{\vbox{\center\LARGE{#1}}\vspace{5mm}}
\renewcommand{\author}[1]{\vbox{\center#1}\vspace{5mm}}
\newcommand{\address}[1]{\vbox{\center\em#1}}

\renewcommand{\date}[1]{\vbox{\center#1}}

